# Structural, elastic, electronic and optical properties of a newly predicted layered-ternary Ti$_4$SiN$_3$: A First-principles study

*M. M. Hossain, M. S. Ali, A. K. M. A. Islam[*]*

*Department of Physics, Rajshahi University, Rajshahi 6205, Bangladesh*

**A B S T R A C T**

We study a newly predicted layered-ternary compound Ti$_4$SiN$_3$ in its α- and β-phases. We calculate their mechanical, electronic and optical properties and then compare these with those of other compounds M$_4$AX$_3$ (M = V, Ti, Ta; A = Si, Al; X = N, C). The results show that the hypothetical Ti$_4$SiN$_3$ shows an improved behavior of the resistance to shape change and uniaxial tensions and a slight elastic anisotropy. The electronic band structures for both α- and β-Ti$_4$SiN$_3$ show metallic conductivity in which Ti $3d$ states dominate. The hybridization peak of Ti $3d$ and N $2s$ lies lower in energy than that of Ti $3d$ and Si $3p$ states which suggests that the Ti $3d$ – N $2s$ bond is stronger than the Ti $3d$ – Si $3p$ bond. Using band structure we discuss the origin of different features of optical properties. The α-phase of predicted compound has improved behavior in reflectivity compared to those of similar types of compounds.

## 1. Introduction

The so-called M$_{n+1}$AX$_n$ phases (where M is a transition metal, A is a group A-element usually belonging to the groups IIIA and IVA, X is C or N, and n = 1-3) have recently attracted more and more attention due to the combination of properties usually associated with metals and ceramics [1-23]. This type of ceramics possesses metal-like properties such as high Young's modulus, good thermal and electrical conductivity, and excellent thermal shock resistance [2, 23]. The materials are exceptionally damage tolerant, remarkably ductile and easily machinable by conventional cutting tools which make them technologically interesting materials. Their unique combination of metallic and ceramic properties makes them suitable for possible applications in a variety of high temperature applications or in other extreme environments.

Currently, more than fifty M$_2$AX (211) compounds, five M$_3$AX$_2$ (312) compounds (Ti$_3$SiC$_2$, Ti$_3$GeC$_2$, Ti$_3$AlC$_2$, Ti$_3$SnC$_2$ and Ta$_3$AlC$_2$) and eight M$_4$AX$_3$ (413) compounds (Ti$_4$AlN$_3$, Ti$_4$SiC$_3$, Ti$_4$GeC$_3$, Ti$_4$GaC$_3$, Ta$_4$AlC$_3$, Nb$_4$AlC$_3$, V$_4$AlC$_3$ and V$_4$AlC$_{2.69}$) have been synthesized [12-20]. The theoretical studies on M$_4$AX$_3$ are mainly focused on Ti$_4$AlN$_3$, Ti$_4$SiC$_3$, Ta$_4$AlC$_3$ and Nb$_4$AlC$_3$. Among these Ti$_4$AlN$_3$ and Ti$_4$SiC$_3$ have drawn a lot of attention recently [22, 23, 27]. Sun [22] and Holm *et al*. [24] investigated the structural, electronic and mechanical properties of Ti$_4$AlN$_3$ using the first-principles method. Their results explained the conductive behavior and bonding character of Ti$_4$AlN$_3$ and predicted its mechanical data including elastic constants, bulk modulus and shear modulus. Li *et al*. [25] calculated the dielectric function and the reflectivity spectrum of Ti$_4$AlN$_3$. They found that Ti$_4$AlN$_3$ could be a good reflecting material for use on spacecraft to avoid solar heating. Magnuson *et al*. [26] investigated the electronic structure and chemical bonding of $Ti_4SiC_3$ and proposed that the analysis of the electronic structure can provide an increased understanding of the physical properties of materials. Based on these studies, it is envisaged that there may be other compounds that could be predicted with a different A-group elements which could show more desirable properties.

In order to explore new field of $M_4AX_3$ phases we predict a new layered-ternary compound $Ti_4SiN_3$ and investigate its stability, elastic, electronic, and optical properties. We will compare the properties of the newly predicted compound with those of similar types of compounds, where available.

## 2. Computational details

Our calculations are carried out by computational methods implemented in CASTEP [28] which uses the plane-wave pseudopotential based on density functional theory (DFT) with the generalized gradient approximation (GGA) in the scheme of Perdew-Burke-Ernzerhof (PBE) [29]. The interactions between ion and electron are represented by ultrasoft Vanderbilt-type pseudopotentials for Ti, Si and N atoms [30]. The elastic constants are calculated by the 'stress-strain' method. The calculations use a plane-wave cutoff energy 500 eV. For the sampling of the Brillouin zone, $13 \times 13 \times 2$ k-point grids generated according to the Monkhorst-Pack scheme [31] are utilized. On the other hand $9 \times 9 \times 2$ k-point grids have been used for the calculation of elastic constants. Geometry optimization is achieved using convergence thresholds of $5 \times 10^{-6}$ eV/atom for the total energy, 0.01 eV/Å for the maximum force, 0.02 GPa for the maximum stress and $5 \times 10^{-4}$ Å for maximum displacement.

## 3. Results and discussion

*3.1 Structural properties*

The predicted layered-ternary compound $Ti_4SiN_3$ is assumed to have a crystal structure similar to $Ti_4AlN_3$ and other $M_4AX_3$ compounds. Further it is also assumed to possess two types of structures i.e. α- and β- forms with two different stacking sequences which is similar to $Ti_4AlN_3$ [24]. We first perform the geometry optimization as a function of normal stress by minimizing the total energy of the proposed compound for both these phases, i.e. α- and β-$Ti_4SiN_3$. The procedure leads to a successful optimization of structures. The optimized crystal structures are shown in Fig. 1. The α-phase ($E$ = -14690.079 eV) is found to be energetically more favorable than the β-phase ($E$ = -14689.554 eV). The optimized cell parameters for both phases of $Ti_4SiN_3$ are given in Table 1, along with those of other compounds.

**Table 1.** The optimized structural parameters for both the phases of $Ti_4SiN_3$.

| Phases | $a$ (Å) | $c$ (Å) | $c/a$ | $V_0$ (Å$^3$) | Ref. |
|---|---|---|---|---|---|
| α-$Ti_4SiN_3$ | 2.9999 | 22.5967 | 7.5325 | 176.11 | This |
| | 3.0 | 22.58 | 7.527 | 175.99 | [48]$^T$ |
| β-$Ti_4SiN_3$ | 2.9830 | 22.8944 | 7.6749 | 176.43 | This |
| | 2.99 | 22.87 | 7.649 | 177.07 | [48]$^T$ |
| $Ti_4AlN_3$ | 2.994 | 23.417 | 7.821 | 181.75 | [23]$^T$ |
| | 2.991 | 23.396 | 7.822 | 181.26 | [15]$^E$ |

T= Theoretical, E= Experimental.

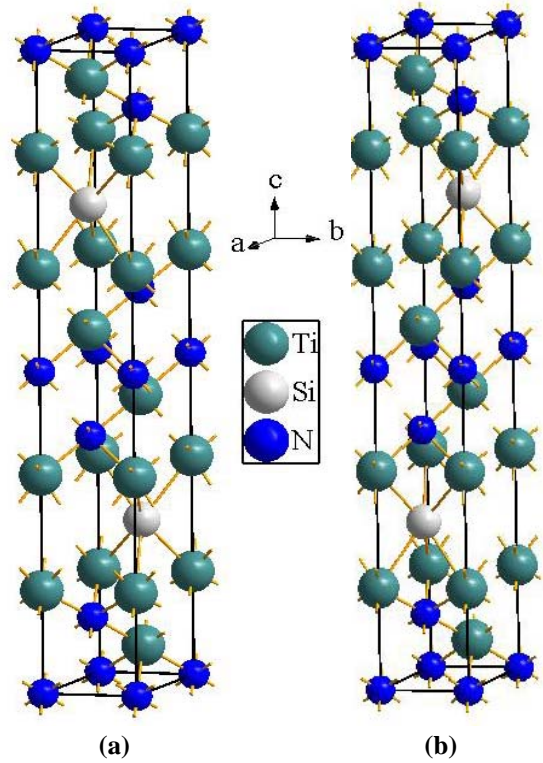

**Fig. 1.** The optimized crystal structures of (a) α-Ti$_4$SiN$_3$ and (b) β-Ti$_4$SiN$_3$.

*3.2. Elastic constants and mechanical stability*

In order to study the mechanical properties of α- and β-Ti$_4$SiN$_3$, we have calculated the elastic constants $C_{ij}$, bulk modulus $B$, shear modulus $G$, Young's modulus $Y$, and Poisson's ratio υ. The calculated results are shown in Table 2 along with other available theoretical and experimental results of several M$_4$AX$_3$ compounds. The well-known Born stability criteria [32] involving elastic constants are: $C_{11} > 0$,  $C_{11}-C_{12} > 0$, $C_{44} > 0$,  $(C_{11} +C_{12}) C_{33} - 2C^2_{13} > 0$. It is seen that the Born stability criteria are satisfied for both α- and β-Ti$_4$SiN$_3$ and hence they are mechanically stable under elastic strain perturbations. There are six elastic constants ($C_{11}$, $C_{12}$, $C_{13}$, $C_{33}$, $C_{44}$, $C_{66}$) for both structures and only five of them are independent, since $C_{66}$ = ($C_{11}$- $C_{12}$) /2. The values of $C_{11}$ for both α- and β-Ti$_4$SiN$_3$ are smaller than those of Ti$_4$AlN$_3$, which show relatively lower resistances against the principal strain $\varepsilon_{11}$. The $C_{44}$-value of α-Ti$_4$SiN$_3$ (40 GPa) is larger than that of Ti$_4$AlN$_3$, which indicates higher resistances to basal and prismatic shear deformations compared to Ti$_4$AlN$_3$.

We know that crystals are usually prepared and investigated as polycrystalline materials in the form of aggregated mixtures of micro-crystallites with a random orientation. It is quite useful to estimate the corresponding parameters for the polycrystalline materials. The theoretical polycrystalline elastic moduli for both structures may be calculated from the set of six elastic constants. Hill [33] proved that the Voigt and Reuss equations represent upper and lower limits of the true polycrystalline constants. He showed that the polycrystalline moduli are the arithmetic mean values of the moduli in the Voigt ($B_V$, $G_V$) and Reuss ($B_R$, $G_R$) approximation, and are thus given by

Hill's bulk modulus, $B_H \equiv B = ½(B_R + B_V)$, where $B_R$ and $B_V$ are the Reuss's and Voigt's bulk modulus respectively. Again Hill's shear modulus, $G_H \equiv G = ½(G_R + G_V)$, where $G_R$ and $G_V$ are the Reuss's and Voigt's shear modulus, respectively. The expression for Reuss and Voigt moduli can be found in Ref. [34]. The polycrystalline Young's modulus $Y$, and Poisson's ratio $v$, are then computed from these values using the relationships [35]: $Y = 9BG/(3B + G)$, $v = (3B − Y)/6B$. We note that the Young's modulus $Y$ of α-Ti$_4$SiN$_3$ phase is larger than that of Ti$_4$AlN$_3$. Therefore, the α-Ti$_4$SiN$_3$ phase compared to Ti$_4$AlN$_3$ shows a better performance of the resistance to shape change and against uniaxial tensions. We also note that the bulk modulus $B$ of α-Ti$_4$SiN$_3$ is higher than that of Ti$_4$AlN$_3$.

The so called shear anisotropy factor, A = $2C_{44}/(C_{11} - C_{12})$ is often used [36] to represent the elastic anisotropy of crystals. The value A = 1 represents completely elastic isotropy, while value smaller or greater than this measures the degree of elastic anisotropy. From Table 2 we see that both α- and β-Ti$_4$SiN$_3$ show completely anisotropic behavior.

**Table 2.** Elastic constsnts $C_{ij}$, the bulk modulus $B$, shear modulus $G$, Young's modulus $Y$ (all in GPa), Poisson's ratio $v$, anisotropic factor $A$, linear compressibility ratio $k_c/k_a$, and ratio $G/B$, at zero pressure.

| Phase | Monocrystal | | | | | | Polycrystal | | | | | | |
|---|---|---|---|---|---|---|---|---|---|---|---|---|---|
| | $C_{11}$ | $C_{12}$ | $C_{13}$ | $C_{33}$ | $C_{44}$ | $C_{66}$ | $B$ | $G$ | $Y$ | $v$ | $A$ | $k_c/k_a$ | $G/B$ |
| α-Ti$_4$SiN$_3$[a] | 387 | 126 | 133 | 392 | 168 | 130 | 217 | 144 | 353 | 0.29 | 1.29 | 0.95 | 0.66 |
| β-Ti$_4$SiN$_3$[a] | 385 | 128 | 119 | 404 | 145 | 128 | 121 | 137 | 339 | 0.23 | 1.13 | 0.96 | 0.64 |
| Ti$_4$AlN$_3$[b] | 420 | 73 | 70 | 380 | 128 | 173 | 183 / 214.97[c] | 144 | 342 | 0.19 | 0.74 | 1.14 | 0.79 |
| α-Ta$_4$SiC$_3$[d] | 396 | 190 | 180 | 391 | 207 | 103 | 254 | 138 | 350 | 0.27 | 2.00 | 1.07 | 0.54 |
| β-Ta$_4$SiC$_3$[d] | 397 | 148 | 190 | 397 | 133 | 124 | 250 | 121 | 312 | 0.29 | 1.06 | 0.80 | 0.48 |
| V$_4$AlC$_3$[e] | 458 | 107 | 110 | 396 | 175 | | 218 | 170 | | | 1.21 | 0.78 |

[a]This work, [b]Ref. [24], [c]Ref. [23], [d]Ref. [39], [e]Ref. [27].

The ratio between linear compressibility coefficients, $k_c/k_a$ can also indicate the elastic anisotropy of materials [37]. For hexagonal crystals, the ratio may be expressed as $k_c/k_a = (C_{11}+C_{12}-2C_{13}) / (C_{33}-C_{13})$. The data obtained for both phases (Table 2) indicates that the compressibility along $c$-axis is smaller than along $a$-axis. One of the most widely used malleability indicators of materials is Pugh's ductility index ($G/B$) [38]. As is known, if $G/B$ < 0.5 the material will have a ductile behavior, whereas if $G/B$ > 0.5 the material is brittle. According to this indicator (Table 2), both the phases of Ti$_4$SiN$_3$ are near the borderline of brittleness.

Finally, the obtained values of the Poisson's ratio $v$ are 0.29 (α-Ti$_4$SiN$_3$) and 0.23 (β-Ti$_4$SiN$_3$). We know that the Poisson's ratio for brittle covalent materials is small, whereas for metallic materials it is typically 0.33 [39]. Thus the materials show the characteristics of being more in the latter category.

*3.3. Electronic properties*

We present the band structure and the densities of states (DOSs) of α- and β-Ti$_4$SiN$_3$ in Figs. 2 and 3, respectively. The valence and conduction bands of both the phases overlap considerably and there are many bands crossing the Fermi level. Thus α- and β-Ti$_4$SiN$_3$ should show metallic conductivity. Moreover, these energy bands around the Fermi level are mainly from the Ti 3$d$ states. This indicates that the Ti 3$d$ states dominate the conductivity of Ti$_4$SiN$_3$. We observe that the lowest three energy bands, i.e. from -17 to -14.5 eV, mainly come from the N 2$s$ states, with little contribution from the Ti 3$d$ states. The energy bands between -7 to 0 eV are dominated by hybridized Ti 3$d$, Si 3$s$/3$p$ and N 2$p$ states. Similar results are found in Ti$_4$AlN$_3$ [22], V$_4$AlC$_3$ [27], α- and β-Ta$_4$SiC$_3$ [39]. The strength of peaks for Si 3$s$/3$p$ is weaker than that for the N 2*p* states. It is seen that the hybridization peak of Ti 3*d* and N 2*s* lies lower in energy than that of Ti 3*d* and Si 3*p* states. This suggests that the Ti 3*d* – N 2*s* bond is stronger than the Ti 3*d* – Si 3*p* bond.

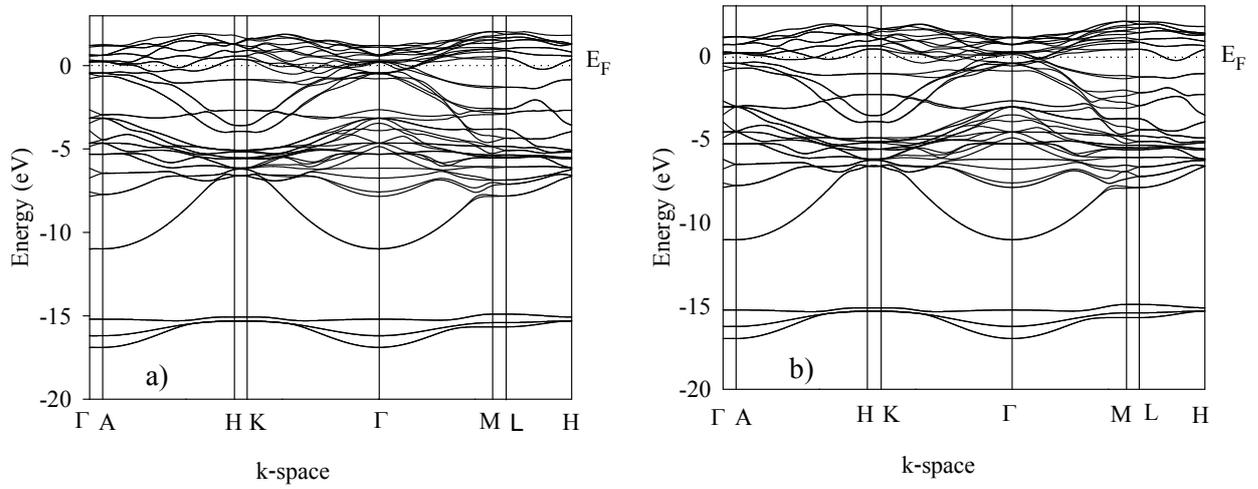

**Fig. 2.** The energy band structures of (a) α-Ti$_4$SiN$_3$ and (b) β-Ti$_4$SiN$_3$.

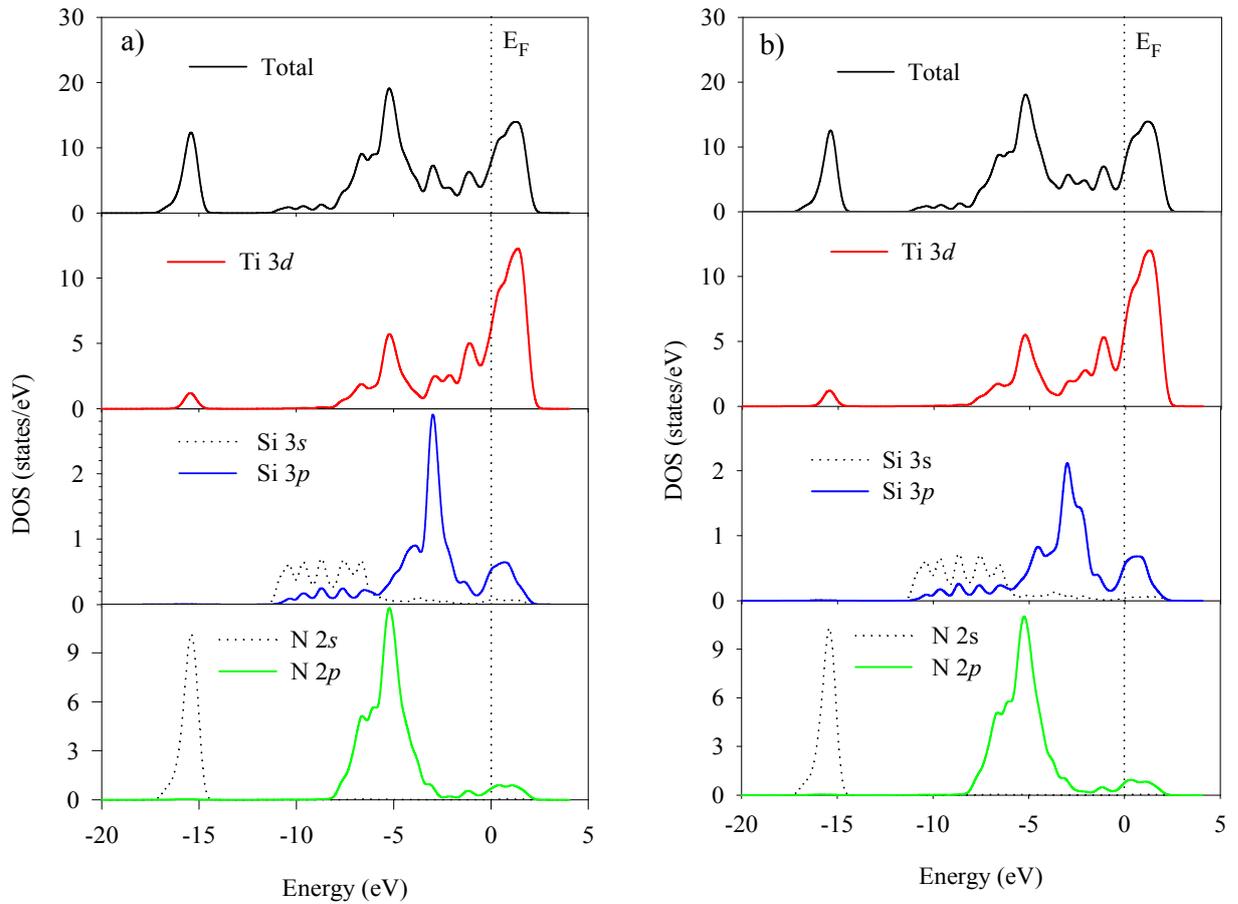

**Fig. 3.** Total and partial DOS of (a) α-Ti$_4$SiN$_3$ and (b) β-Ti$_4$SiN$_3$.

*3.4. Optical properties*

In our calculation of optical properties, we used a Gaussian smearing of 0.5 eV to smear out the Fermi level so that the k-points will be more effective on the Fermi surface. The reflectivity of $Ti_4SiN_3$ as a function of photon energy is presented in Fig. 4 along with the spectra of α-$Nb_4SiC_3$, $Ti_4AlN_3$ and $V_4AlC_3$. We observe that the reflectivity value of α-$Ti_4SiN_3$ is always higher than those of $Ti_4AlN_3$, $V_4AlC_3$ and α-$Nb_4SiC_3$ till 5 eV photon energy. Therefore, the capability of the predicted α-$Ti_4SiN_3$ to reflect solar light is stronger than the other existing $Ti_4AlN_3$, $V_4AlC_3$, and α-$Nb_4SiC_3$. The β-phase of $Ti_4SiN_3$ exhibits more or less similar behavior as of $V_4AlC_3$, and α-$Nb_4SiC_3$. We can see that the spectrum of α-$Ti_4SiN_3$ is roughly nonselective with reflectivity 48-68%.

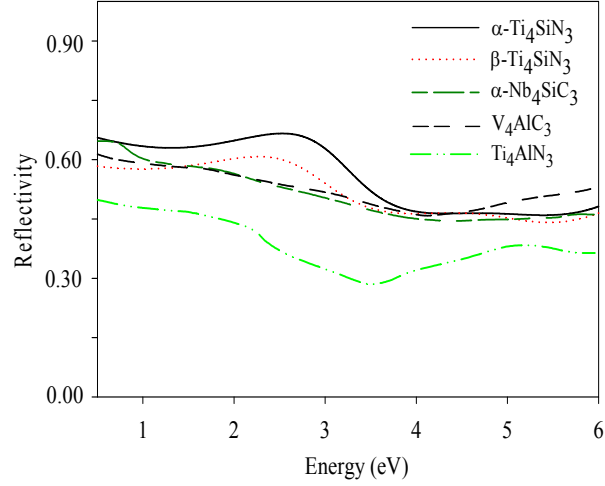

**Fig. 4.** Calculated reflectivity spectra of α- and β-$Ti_4SiN_3$ to those of α-$Nb_4SiC_3$ [40], $V_4AlC_3$ [27] and $Ti_4AlN_3$ [25].

Figs. 5 and 6 show the optical functions of α- and β-$Ti_4SiN_3$ calculated for photon energies up to 20 eV. In Fig. 5a, the α-$Ti_4SiN_3$ has an absorption band in the low energy range due to its metallic nature. Its absorption spectrum rises sharply and has two peaks at ~ 2.5 and ~ 7 eV. It then decreases rapidly to zero at ~12 eV. There is also a smaller peak at ~16.7 eV. The first two peaks are associated with the transition from the Si/N *p* to the Ti *d* states. Nearly the same features can be seen for β-$Ti_4SiN_3$. The energy loss of a fast electron traversing in the material is manifested in the energy-loss spectrum [41]. Its peak is defined as the bulk plasma frequency $\omega_p$, which occurs when $\varepsilon_1$ reaches the zero point [42, 43].

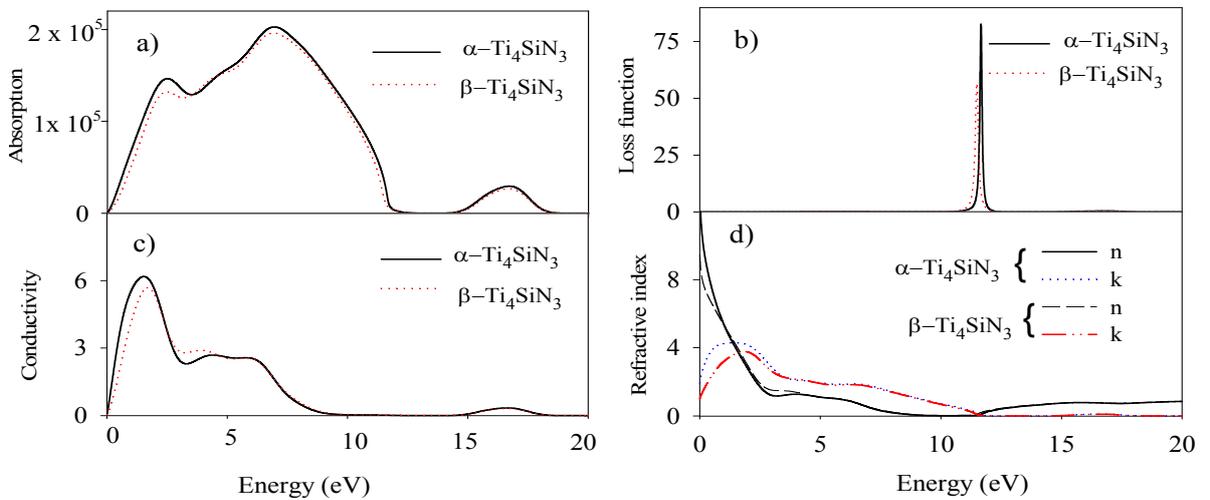

**Fig. 5.** (a) Absorption spectrum, (b) photo-conductivity, (c) energy-loss spectrum and (d) refractive index of $Ti_4SiN_3$.

In the energy-loss spectrum (Fig. 5b) we observe that the plasma frequency $\omega_p$ for α-$Ti_4SiN_3$ is 11.5 eV (11.6 eV for β-$Ti_4SiN_3$). Thus when the frequency of incident light is higher than ~11.5 eV (~ 11.6 eV for β-phase), the material becomes transparent. Since the materials have no band gap as evident from band structures, the photoconductivity for both the phases starts when the photon energy is zero (Fig. 5c). Photoconductivity and hence electrical conductivity of a material increases as a result of absorbing photons [44]. The phase of α-$Ti_4SiN_3$ presents three peaks at 2, 4 and 6 eV similar to three peaks at 2.5, 3.8 and 6 eV for β-$Ti_4SiN_3$. The real part (refractive index, n) and imaginary part (extinction coefficient, k) of complex refractive index have been shown in Fig. 5d. For both phases the real part of refractive index, $n$ is zero at about 9 eV whereas for imaginary part of refractive index, $k$ is zero at about 12 eV.

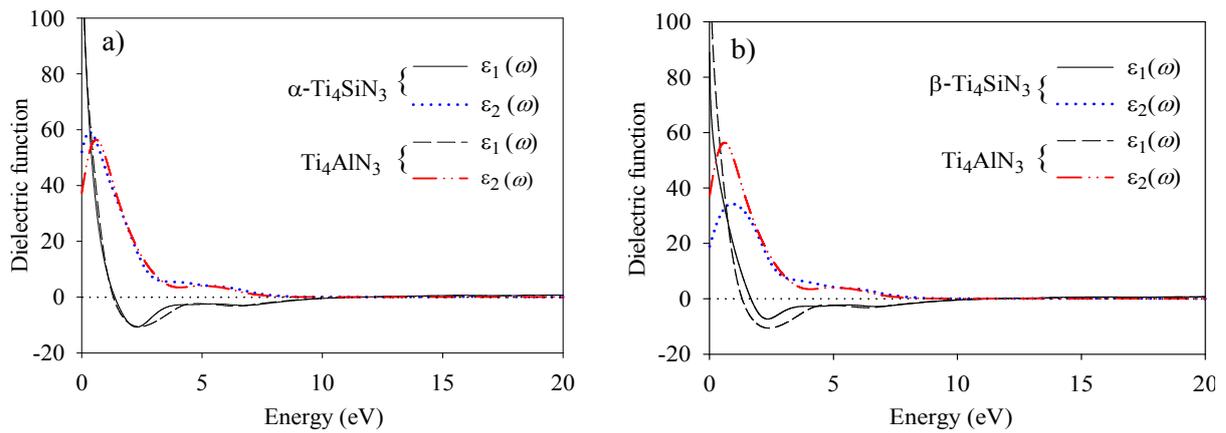

**Fig. 6.** The calculated dielectric functions of (a) α-$Ti_4SiN_3$ and (b) β-$Ti_4SiN_3$ in comparison to those of $Ti_4AlN_3$.

The imaginary and real parts of the dielectric function for both the phases are displayed in Fig. 6 in comparison to those of $Ti_4AlN_3$. It is observed that the real part $\varepsilon_1$ vanishes at about 11.5 eV. This corresponds to the energy at which the reflectivity exhibits a sharp drop at around 11.6 eV and the energy-loss function (Fig. 5b) also shows the first peak [45]. This peak in energy-loss function at about 11.5 eV which arises as $\varepsilon_1$ goes through zero and $\varepsilon_2$ is small at such energy. Thus this fulfills the condition for plasma resonance at 11.5 eV ($\hbar\omega_p$= 11.5 eV). We also observe that the calculated dielectric functions of α- and β-$Ti_4SiN_3$ are nearly similar to those of $Ti_4AlN_3$. The peak for < 1eV for imaginary part $\varepsilon_2$ is due to transitions within the Ti 3$d$ bands. The large negative values of $\varepsilon_1$ indicate that both phases of $Ti_4SiN_3$ show Drude-like behavior.

## 4. Conclusions

The structural stability and mechanical, electronic and optical properties of the newly predicted layered-ternary α- and β-$Ti_4SiN_3$ are investigated. Both the structures are found to be mechanically stable. The elastic constants, bulk modulus, shear modulus and Young's modulus of α- and β-$Ti_4SiN_3$ are compared to those of other similar $M_4AX_3$ compounds. The predicted $Ti_4SiN_3$ compound shows an improved behavior of the resistance to shape change and uniaxial tensions and a slight elastic anisotropy. The electronic band structures for both phases show metallic conductivity. Moreover, the Ti-N bonding is stronger than the Ti-Si bonding in $Ti_4SiN_3$ indicating that Ti-N bond is more resistant to deformation than the Ti-Si bond. Using the band structure, we have discussed the origin of the features that appear in the optical properties. The reflectivity spectrum shows that the predicted compound $Ti_4SiN_3$ (particularly the α-phase) is a better candidate material as a coating to avoid solar heating than the other existing $Ti_4AlN_3$, $V_4AlC_3$ and α-$Nb_4SiC_3$ compounds. The study should provide incentives for further experimental investigation which would pave the way for practical application for Ti$_4$SiN$_3$.